\documentclass[aps,prd,twocolumn,amssymb,amsmath,showpacs,a4paper,superscriptaddress]{revtex4-1}

\usepackage{graphicx}
\usepackage{amsfonts}
\usepackage{amsmath}
\usepackage{hyperref}
\usepackage{dcolumn}
\usepackage{bm}

\usepackage{latexsym}
\usepackage{amsfonts}
\usepackage{psfrag}
\usepackage{graphicx}
\usepackage{amssymb,amsmath} 
\usepackage{appendix}

\usepackage{color}
\definecolor{myOrange}{rgb}{1,0.5,0.1}
\definecolor{myRed}{rgb}{0.8,0.1,0.1}
\definecolor{myGreen}{rgb}{0.7,0,0.8}
\definecolor{myGray}{rgb}{0.6,0.6,0.6}

\definecolor{light-gray}{gray}{0.95}

\begin{document}

\setlength{\pdfpagewidth}{8.5in}
\setlength{\pdfpageheight}{11in}

\title{Dynamics Landscape for Acoustic Superradiance}

\author{Cisco Gooding}
\affiliation{
	School of Mathematical Sciences, 
	University of Nottingham, UK
	}
\date{\today}

\begin{abstract}
We analyze the behaviour of acoustic vortex beams interacting with rotating, fluid-saturated porous materials. Regions of the parameter space that exhibit distinct dynamical features are identified, with a focus on features that are relevant to the characterization of rotational superradiance. We discuss the similarities and differences between two recent proposals to observe acoustic superradiance with rotating, air-saturated sound absorbers. Finally, theoretical predictions for macroscopic acoustic scattering, obtained by averaging over interactions between the fluid and the porous material at the microscopic level, are compared with predictions of the first-Born approximation. 
\end{abstract}

\maketitle

\section{Introduction}

The theoretical study of rotational superradiance began in the early $1970$s in the context of electromagnetic \cite{Zeldovich71,Zeldovich72} and gravitational systems \cite{Penrose71,Bekenstein73}, and was later applied also to fluid dynamical systems \cite{Silke2016}. The first experimental confirmation of rotational superradiance took place not long ago, using surface water waves on a vortex flow \cite{Silke2017}, which created a renewed interest in superradiance research.

More recently, a new scattering configuration was proposed to superradiate acoustic orbital angular momentum (OAM) beams \cite{GWU2018}. The new arrangement involves scattering the OAM beam along the symmetry axis of a rotating sound absorber, rather than perpendicular to this axis, as the incident mode is in the standard formulation \cite{Zeldovich71,Zeldovich72}. 

In this work, we examine the landscape of parameter regimes for acoustic superradiance in this new configuration, by extending the analysis of \cite{GWU2018} to allow transmission of the incident OAM beam through the rotating absorber. Interpreted appropriately, this extended analysis can also be applied to a related proposal to observe superradiant gain in acoustic systems \cite{Faccio}. We compare the model presented here to the theoretical description offered in \cite{Faccio}, with a focus on approximations used and their domains of applicability.

\section{Theory}

The fluid dynamical model we consider is based on the model described in \cite{GWU2018}; we refer the reader to that work for further details. Our (barotropic) fluid is contained inside a cylindrical tube of radius $R$. Two pure fluid regions are separated by a rotating, sound absorbing disk of finite thickness that is coaxial with the tube. On one side of the disk (which we call the incident side), an acoustic OAM beam is generated and directed through the tube towards the disk. The fluid on the incident side is otherwise motionless, as is the fluid on the other side of the disk (which we call the transmitted side), with background density $\rho_0$. 

Acoustic OAM modes within the pure fluid  are defined by a potential $\Phi$, and the corresponding fluid velocity is $\bm{u}=\nabla \Phi$. We denote the wave frequency by $\omega$, the topological charge of the mode by $m$, the sound speed by $c$, and the acoustic pressure by $p=c^2 \rho=i\omega\rho_0\Phi$.

The sound absorbing disk is treated as a porous material, with a characteristic pore scale much smaller than the acoustic wavelength. In anisotropic rigid-framed porous media, quasi-stationary fluids obeys Darcy's law, 
\begin{equation}\label{DarcyLett}
\bm{u}=-\frac{1}{\mu}\bm{K}(\omega)\cdot\nabla p,
\end{equation}
where $\mu$ is the fluid viscosity and $\bm{K}(\omega)$ is the permeability tensor~\cite{ConvectionPorousMedia,DarcyOG,DynamPorousFluids}. The fluid velocity $\bm{u}$ appearing in (\ref{DarcyLett}) is a macroscopic quantity, as Darcy's law can be interpreted as an effective large-scale behaviour obtained by averaging over microscopic structure at the pore scale \cite{Homogenization}. As discussed in \cite{GWU2018}, the analysis of Auriault \cite{AcoustRotDeformPorous} provides a basis for describing acoustics within the rotating absorbing disk, and we will again make use of this approach here.

Perturbations in the fluid within the rotating porous medium obey a generalized Darcy law $\nabla p=-\mu\bm{K}^{-1}(\bar{\omega},\bm{\Omega})\bm{u}$, characterized by the tensor
\begin{equation}\label{ExplicitH}
\bm{K}^{-1}(\bar{\omega},\bm{\Omega})=\frac{1}{K_0}\bm{\delta}-\frac{i\bar{\omega}\rho_0}{\varphi\mu}\bm{A}(\bar{\omega},\bm{\Omega}),
\end{equation}
where $\bm{\Omega}=\Omega\hat{\bm{z}}$ is the angular velocity of the disk, $\bar{\omega}=\omega-m\Omega$ is the convective frequency, $\varphi$ is the porosity of the material, and $\bm{A}$ has components
\begin{equation}
A_{ij}=\left(1+\frac{\Omega^2}{\bar{\omega}^2}\right)\delta_{ij}-\frac{1}{\bar{\omega}^2}\Omega_i\Omega_j-\frac{2i}{\bar{\omega}}\epsilon_{ijn}\Omega_{n}.\label{Inertial}
\end{equation}  
The real part of $\bm{K}^{-1}$ describes dissipation, and the imaginary part describes inertial effects. In the high frequency regime, one can neglect viscous effects, and at leading order one finds
$\bm{K}^{-1}=\frac{-i\bar{\omega}\rho_0}{\varphi\mu}\bm{A}$.  For a macroscopically isotropic medium in the low frequency regime, neglecting the inertial terms leads to $\bm{K}^{-1}=\frac{1}{K_0}\bm{\delta}$, where $\bm{\delta}$ is the unit tensor and $K_0$ is the static permeability \cite{GWU2018}.

Acoustic waves within the porous medium can be described by a single dynamical equation for the pressure,
\begin{equation}\label{PressDarcy}
\beta\left(\frac{\bar{\omega}}{c}\right)^2 p = \frac{i\bar{\omega} \rho_0}{\mu\varphi}\nabla\cdot\left[\bm{K}\nabla p\right],
\end{equation}
with the compressibility $\beta$ being a complex function of the frequency $\omega$. The tensorial nature of the permeability $\bm{K}$ appearing in (\ref{PressDarcy}) is induced by rotation, since we are assuming that the material structure is otherwise isotropic. If one assumes that $\bm{K}$ is just a scalar times the unit tensor, as has been done recently for a related proposal \cite{Faccio}, then inertial contributions to the dynamics (\ref{PressDarcy}) will have discrepancies of order $\Omega/\bar{\omega}$. These discrepancies can become irrelevant for wave frequencies and rotation rates well below the viscous frequency scale $\bar{\omega}_c$ \cite{Faccio}, though for rapid rotation rates one must take into account the full dynamics described by (\ref{PressDarcy}).

Another tensor of interest is the tortuosity $\bm{\alpha}(\bar{\omega})=(i\varphi \mu/\bar{\omega} \rho_0)\bm{K}^{-1}$ \cite{Porous}; assuming isotropy in the transverse plane, we can use this tensor to write the pressure equation (\ref{PressDarcy}) as
\begin{equation}\label{TortPressDiff}
-\left(\frac{\bar{\omega}}{c}\right)^2 p=\alpha^{-1}_{rr}\left(\frac{1}{r}\partial_r\left(r\partial_r p\right) + \partial_{\bar{\theta}}^2 p\right)+\alpha^{-1}_{zz} \partial_z^2 p,
\end{equation}
where $\bar{\theta}$ is the azimuthal angle with respect to a coordinate system rotating with the disk, and we have taken $\beta$ to be unity \cite{GWU2018}. The permeability tensor $\bm{K}$ is proportional to the inverse tortuosity tensor via $\bm{K}=(i\varphi \mu/\bar{\omega} \rho_0)\bm{\alpha}^{-1}$, and can be obtained by inverting the matrix equation produced by expressing (\ref{ExplicitH}) in cylindrical coordinate components. Defining $\bar{\omega}_c=\mu\varphi/\rho_0 K_0$ and $\bar{\omega}_\Omega=\bar{\omega}(1+\Omega^2/\bar{\omega}^2)$, the independent nontrivial components of $\bm{\alpha}^{-1}$ can be written
\begin{equation}\label{AlphaInvzz}
\alpha_{zz}^{-1}=\frac{\bar{\omega}}{\bar{\omega}+i\bar{\omega}_c}
\end{equation}
and
\begin{equation}\label{AlphaInvrr}
\alpha_{rr}^{-1}=\frac{\bar{\omega}\left(\bar{\omega}_\Omega+i\bar{\omega}_c\right)}{\left(\bar{\omega}_\Omega+i\bar{\omega}_c\right)^2-4\Omega^2}.
\end{equation}

\subsection{Piecewise Solutions}

The pure fluid region within the cylindrical tube on the incident side of the rotating absorber is defined by $z<0$, with $z=0$ being the first fluid-disk interface. The acoustic potential $\Phi$ is given by solutions to the Helmholtz equation of the form
\begin{equation}\label{TotalPotSup}
\Phi=J_m(k_r r)e^{im\theta}\left(\alpha^+ e^{ik_z z}+\alpha^-e^{-ik_z z}\right)e^{-i\omega t}.
\end{equation}
Here $\alpha^+$ is the incident mode amplitude, $\alpha^-$ is the reflected mode amplitude, $k_r$ is the radial wavenumber, $k_z$ is the axial wavenumber, and $J_m$ is a Bessel function of the first kind.

The disk region is defined by $0<z<l$. In this region, the pressure can be written as
\begin{equation}
p = J_{\bar{m}}(k_r r)e^{i m\bar{\theta}}\left(\alpha_+e^{i\bar{k}_z z}+\alpha_-e^{-i\bar{k}_z z}\right)e^{-i\bar{\omega}t}.
\end{equation}
Inserting this ansatz into (\ref{TortPressDiff}), one finds the dispersion relation
\begin{equation}\label{DispersionOm}
\left(\frac{\bar{\omega}}{c}\right)^2 =\alpha^{-1}_{rr}\bar{k}_r^2 +\alpha^{-1}_{zz} \bar{k}_z^2.
\end{equation}

Finally, the pure fluid region on the transmitted side is defined by $z>l$. The transmitted acoustic fluid velocity field is $\tilde{\bm{u}}=\nabla \tilde{\Phi}$, and the acoustic pressure is $\tilde{p} = i \omega \rho_0 \tilde{\Phi}$, with the potential $\tilde{\Phi}$ given by Helmholtz solutions of the form
\begin{equation}\label{TotalPotSup}
\tilde{\Phi}=J_m(k_r r)e^{im\theta}\tilde{\alpha}^+ e^{ik_z z}e^{-i\omega t}.
\end{equation}
Here $\tilde{\alpha}^+$ is the transmitted mode amplitude.

\subsection{Boundary Conditions}

The behaviour of the fluid at each type of boundary was discussed in \cite{GWU2018}; we briefly describe how they are applied to the transmitting case presented here. At the surface of the tube, $r=R$, we enforce the impermeability boundary condition
\begin{equation}\label{BCR}
\bm{u}\cdot\hat{\bm{r}}|_{r=R}=0,
\end{equation}
which constrains the radial wavenumber $k_r$ to satisfy
\begin{equation}\label{RadCon}
J_m'(k_r R) = 0.
\end{equation}

Pressure continuity at the $z=0$ and $z=l$ fluid-disk interfaces yields
\begin{equation}
i \omega \rho_0\left(\alpha^++\alpha^-\right)=\alpha_++\alpha_-
\end{equation}
and
\begin{equation}
i \omega \rho_0\tilde{\alpha}^+e^{ik_z l}=\alpha_+e^{i\bar{k}_z l}+\alpha_-e^{-i\bar{k}_z l},
\end{equation}
respectively. Introducing the definition
\begin{equation}
\chi \equiv \frac{i\varphi\omega^2 \bar{k}_z \alpha^{-1}_{zz}}{\bar{\omega}^2 k_z},
\end{equation}
continuity of normal fluid displacement can be expressed at $z=0$ as
\begin{equation}
\omega\rho_0\left(\alpha^+-\alpha^-\right)=-\chi\left(\alpha_+-\alpha_-\right),
\end{equation}
and at $z=l$ as
\begin{equation}
\omega\rho_0\tilde{\alpha}^+e^{ik_z l}=-\chi\left(\alpha_+ e^{i\bar{k}_z l}-\alpha_- e^{-i\bar{k}_z l}\right).
\end{equation}

The boundary conditions allow determination of the transmission amplitude $T=\tilde{\alpha}^+/\alpha^+$ such that
\begin{equation}\label{tran}
T = \frac{2e^{-ik_z l}}{2\cos\left(\bar{k}_z l\right)+\left(\frac{1}{\chi}-\chi\right)\sin\left(\bar{k}_z l\right)},
\end{equation}
with the transmitted amplification factor given by $|T|^2-1$. The reflection amplitude $R=\alpha^-/\alpha^+$ can then be written as
\begin{equation}\label{refl}
R = \frac{\left(\frac{1}{\chi}+\chi\right) \sin\left(\bar{k}_z l\right)}{2\cos\left(\bar{k}_z l\right)+\left(\frac{1}{\chi}-\chi\right)\sin\left(\bar{k}_z l\right)},
\end{equation}
with the reflected amplification factor given by $|R|^2-1$. For completeness, we note that the forward-moving amplitude is given by
\begin{equation}
\frac{\alpha_+}{\alpha^+}=\frac{-\omega\rho_0 e^{-i\bar{k}_z l}\left(1-i\chi\right)}{\chi\left[2\cos\left(\bar{k}_z l\right)+\left(\frac{1}{\chi}-\chi\right)\sin\left(\bar{k}_z l\right)\right]},
\end{equation}
and that the reverse-moving amplitude is given by
\begin{equation}
\frac{\alpha_-}{\alpha^+}=\frac{\omega\rho_0 e^{i\bar{k}_z l}\left(1+i\chi\right)}{\chi\left[2\cos\left(\bar{k}_z l\right)+\left(\frac{1}{\chi}-\chi\right)\sin\left(\bar{k}_z l\right)\right]}.
\end{equation}

\subsection{Thick Disk Limit}

As $l\rightarrow\infty$, the reflection and transmission amplitudes behave as
\begin{equation}\label{Reflim}
R \rightarrow R_\infty\equiv\frac{1+i\chi}{1-i\chi}
\end{equation}
and
\begin{equation}
T\rightarrow 0,
\end{equation}
respectively. The limiting reflection amplitude (\ref{Reflim}) coincides with the thick disk limit of the reflection amplitude derived using a rigid backing \cite{GWU2018} instead of the transmitting case considered here. The quantity (\ref{Reflim}) has modulus-squared
\begin{equation}
|R_\infty|^2 = \frac{1+|\chi|^2-2\text{Im}[\chi]}{1+|\chi|^2+2\text{Im}[\chi]},
\end{equation}
which is less than (greater than) unity for positive (negative) values of $\text{Im}[\chi]$.

\begin{figure*}[t!]
\includegraphics[width=\textwidth]{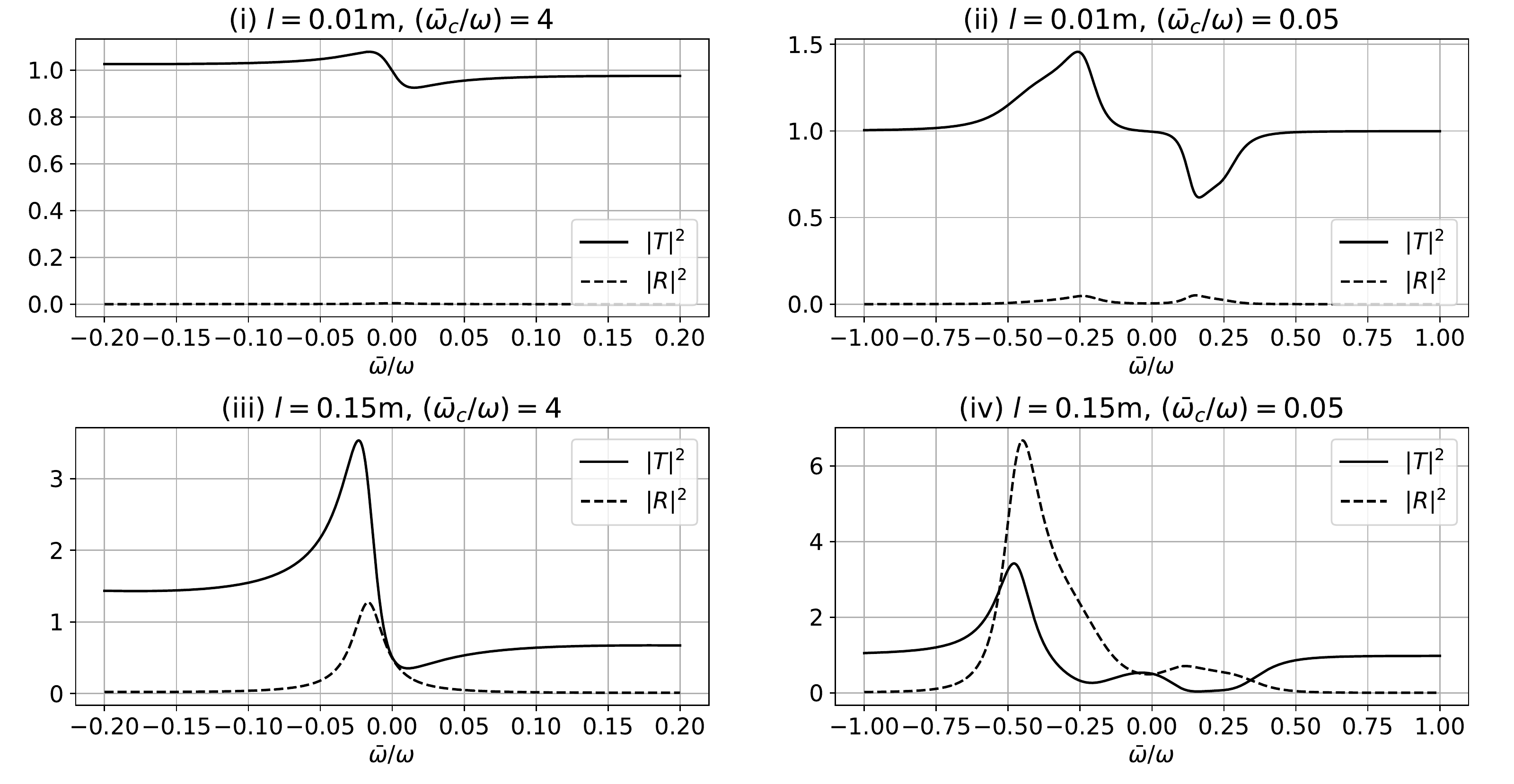} %
\caption{Plot of the scattering behaviour described by equations (\ref{tran}) and (\ref{refl}), using the following parameters: $\omega=2\pi\cdot 100\text{rads}/\text{s}$, $c=343\text{m}/\text{s}$, $k_r = 0.7(\omega/c)$, $m=4$, and $\varphi=0.75$.}
	\label{fig:1}
\end{figure*}

\section{Comparison with Alternative Approach}

As mentioned above in the context of the tensorial nature of the permeability, the related proposal \cite{Faccio} involves low acoustic frequencies and rotation rates. Expressed in terms of the acoustic pressure, the internal equation of motion used for this regime is
\begin{equation}\label{FaccioDyn}
\left(\frac{\partial}{\partial t}+\Omega\frac{\partial}{\partial \theta}\right)^2 p -\Gamma'\nabla^2\left(\frac{\partial}{\partial t}+\Omega\frac{\partial}{\partial \theta}\right)p-c^2\nabla^2 p = 0,
\end{equation}
with corresponding dispersion relation
\begin{equation}\label{ScalarAbsorb}
\bar{k}_z^2 = \frac{1}{\left(1-\frac{i\Gamma' \bar{\omega}}{c^2}\right)}\left(\frac{\bar{\omega}}{c^2}\right)^2-k_r^2.
\end{equation} 
As can be deduced by comparing the form of the dispersion relation (\ref{ScalarAbsorb}) with the general case (\ref{DispersionOm}), the dynamics (\ref{FaccioDyn}) assumes a scalar tortuosity equal to
\begin{equation}\label{ScalarTort}
\alpha_{zz}=\frac{1}{1-\frac{i\Gamma' \bar{\omega}}{c^2}}=\frac{1}{1-\frac{i\eta\bar{\omega}}{\omega}},
\end{equation}
where $\eta\equiv \omega\Gamma'/c^2$. 

In the limit where the disk is much thinner than the acoustic wavelength, the authors of \cite{Faccio} use the ``first-Born'' approximation to obtain a transmission amplitude with modulus-squared given by 
\begin{equation}\label{Tfac}
|T|_{FB}^2\approx 1-2 a l,
\end{equation}
where
\begin{equation}\label{AbsFacc}
a = \frac{\bar{\omega}\omega^2\Gamma'}{2 k_z c^4}.
\end{equation}
We restrict our attention to propagating incident modes, which corresponds to taking the axial wavenumber $k_z$ to be real; the extension to evanescent incident modes is straightforward, and parallels the analysis of \cite{GWU2018}.

The modulus-squared of the scattering amplitudes (\ref{tran}) and (\ref{refl}) are displayed in Figure \ref{fig:1}, for propagating incident modes. Mode amplification occurs when either $|T|^2$ or $|R|^2$ reach above unity. One can observe from these plots that superradiant amplification increases as one transitions between the low frequency regime ($\bar{\omega}_c/\omega=4$) and the high frequency regime ($\bar{\omega}_c/\omega=0.05$). Moreover, one can also observe that transmitted mode amplification dominates for thin disks, whereas reflected mode amplification becomes increasingly prominent as the disk thickness increases. 

Of course, the use of propagating incident modes involves some caveats \cite{GWU2018}. The radial wavenumber constraint (\ref{RadCon}) implies that the axial wavenumber $k_z$ can be written as
\begin{equation}
k_z=\left(\frac{1}{c}\right)\sqrt{\omega^2-\left(\frac{c x_{mn}}{R}\right)^2},
\end{equation}
where $x_{mn}$ is a zero of $J_m'(x)$. When the superradiance condition $\bar{\omega}<0$ is met, $k_z$ is only real when
\begin{equation}\label{Shock}
R\Omega>c\left(\frac{x_{mn}}{m}\right).
\end{equation}
Since only co-rotating modes can satisfy the superradiance condition, we must have $m>0$, which implies that the Bessel zeros satisfy $x_{mn}/m>1$. Then, the inequality (\ref{Shock}) yields $R\Omega > c$, which means that the outer edge of the disk must be moving faster than the sound speed. It is therefore not clear whether the acoustic waveguide described in \cite{Faccio} can fit a suitable propagating OAM mode into the disk at low rotation rates.

The transmission amplitude (\ref{tran}) implies the asymptotic behaviour 
\begin{equation}\label{Tasym}
|T|^2\sim 1 - l\cdot\text{Re}\left[\bar{k}_z\left(\frac{1}{\chi}-\chi\right)\right]+\mathcal{O}(l^2).
\end{equation}
Explicitly, one finds
\begin{equation}\label{asympT}
\text{Re}\left[\bar{k}_z\left(\frac{1}{\chi}-\chi\right)\right]=\text{Im}\left[\frac{\bar{\omega}^2 k_z}{\varphi\omega^2\alpha^{-1}_{zz}}+\frac{\varphi\omega^2\bar{k}_z^2\alpha^{-1}_{zz}}{\bar{\omega}^2 k_z}\right].
\end{equation}

If we take a scalar tortuosity given by (\ref{ScalarTort}), such that
\begin{equation}\label{ScalarDisp}
\bar{k}_z^2=\alpha_{zz}\left(\frac{\bar{\omega}}{c}\right)^2-k_r^2,
\end{equation}
then (\ref{asympT}) can be re-expressed as
\begin{equation}\label{asympTscal}
\text{Re}\left[\bar{k}_z\left(\frac{1}{\chi}-\chi\right)\right]=\frac{\bar{\omega}\eta}{\omega k_z}\left(\frac{\bar{\omega}^2 k_z^2}{\varphi\omega^2\left(1+\frac{\bar{\omega}^2\eta^2}{\omega^2}\right)}+\frac{\varphi\omega^2 k_r^2}{\bar{\omega}^2}\right).
\end{equation}

One can observe from (\ref{asympTscal}) that superradiance occurs for $\bar{\omega}<0$, which coincides with the standard superradiance condition \cite{Zeldovich71,Zeldovich72}. The divergence at the crossover point $\bar{\omega}=0$ signals a breakdown of the scalar tortuosity approximation; even for low sound frequencies and rotation rates, the contributions to the dynamics (\ref{PressDarcy}) that depend on $\Omega/\bar{\omega}$ become significant near $\bar{\omega}=0$. Using the full dispersion (\ref{DispersionOm}), the offending term $\bar{k}_z \chi$ is given by
\begin{equation}
\bar{k}_z \chi=\frac{i\varphi\omega^2\bar{k}_z^2 \alpha^{-1}_{zz}}{\bar{\omega}^2 k_z}=\frac{i\varphi\omega^2}{k_z}\left(\frac{1}{c^2}-k_r^2\frac{\alpha^{-1}_{rr}}{\bar{\omega}^2}\right),
\end{equation}
which remains finite as $\bar{\omega}\rightarrow 0$, since $\alpha^{-1}_{rr}/\bar{\omega}^2\rightarrow 1/\Omega^2$.

The experimental parameters mentioned in \cite{Faccio} are $m=4$, $\eta=10$, $\omega=2\pi\cdot 100\hspace{1pt}\text{rads}/\text{s}$, $k_r=0.7 (\omega/c)$, $c=343\hspace{1pt}\text{m}/\text{s}$, and $l=1\hspace{1pt}\text{cm}$. The crossover point $\bar{\omega}=0$ occurs for $\Omega=2\pi\cdot 25\hspace{1pt}\text{rads}/\text{s}$, and at $\Omega=2\pi\cdot 35\hspace{1pt}\text{rads}/\text{s}$, the first-Born approximation (\ref{Tfac}) predicts a gain of $10\text{\%}$. For a low porosity of $\varphi=0.5$, (\ref{Tasym}) implies a $16\text{\%}$ gain, while for a high porosity of $\varphi=0.9$, (\ref{Tasym}) implies a $28\text{\%}$ gain. 

These predictions overestimate the gain produced by superradiance, due to the $\bar{\omega}^2$ factor in the denominator of the $\bar{k}_z\chi$ term on the right-hand-side of (\ref{Tasym}). Though this overestimation originates from the assumption of a scalar tortuosity (\ref{ScalarTort}), the issue does not manifest itself in the first-Born approximation \cite{Faccio}, which is insensitive to specific features of the boundary value problem. 

To make further contact with the analysis of \cite{Faccio}, one would ideally want to compare the predictions of the first-Born approximation with the scattering solution (\ref{tran}), using the full dispersion (\ref{DispersionOm}). However, since the real part of $\alpha_{zz}$ is unity, and the real part of the $(\bar{\omega}/c)^2$ prefactor appearing in (\ref{ScalarAbsorb}) is $\left(1+(\Gamma')^2\bar{\omega}^2/c^4\right)^{-1}$, connecting the two approaches is not completely straightforward. The real part of the $(\bar{\omega}/c)^2$ prefactor from \cite{Faccio} controls the effective speed of sound within the porous material; likewise, this sound speed is determined in the general case by the real part of the product of the compressibility and the tortuosity. 

\begin{figure}[t]
\includegraphics[width=0.47\textwidth]{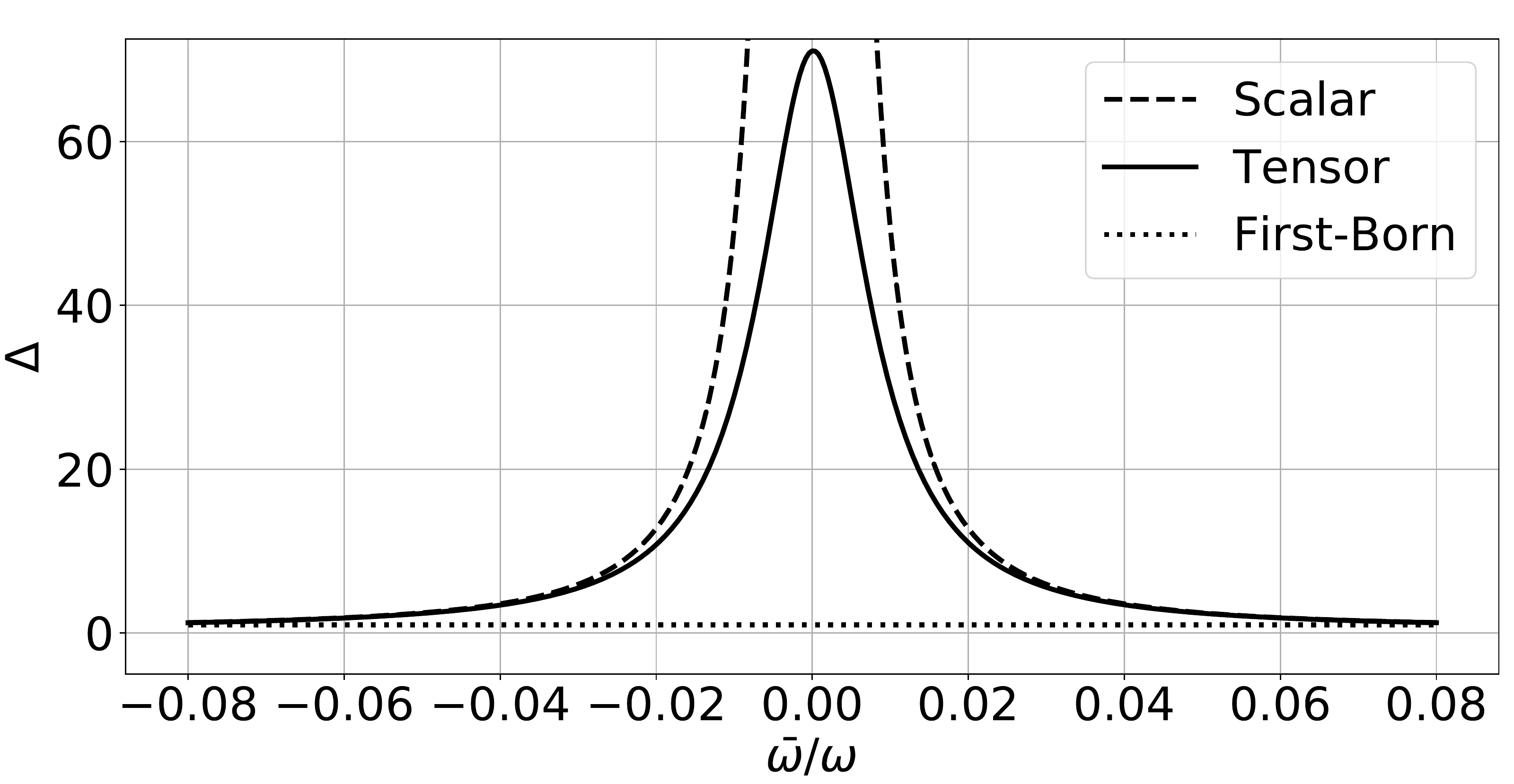} %
\caption{Plot of the $\Delta$ factor appearing in equation (\ref{TensAmp}), using the following parameters: $\omega=2\pi\cdot 100\text{rads}/\text{s}$, $c=343\text{m}/\text{s}$, $k_r = 0.7(\omega/c)$, $m=4$, $\varphi=0.75$, and $\hat{\eta}\equiv\omega\hat{\Gamma}/c^2\equiv\bar{\omega}_c/\varphi\omega=10$. The scalar tortuosity prediction based on (\ref{ScalarDisp}) with $\alpha^{-1}_{zz}$ given by (\ref{AlphaInvzz}) (dashed line), the tensor tortuosity prediction determined by (\ref{delta}) (solid line), and the prediction based on the first-Born approximation (\ref{Tfac}) (dotted line at $\Delta=1$).}
	\label{fig:2}
\end{figure}

The reason this difference between the approaches occurs is that each approach is tailored for a difference frequency regime. For sufficiently low acoustic frequencies, the interaction between sound and the porous material is isothermal, whereas for high enough frequencies the interaction is adiabatic. The behaviour of the scalar tortuosity and the dynamic bulk modulus (which is related to the compressibility, $\beta$) in the high and low frequency regimes was analyzed in \cite{Compress,TortFreq}. In particular, in \cite{TortFreq} the transition between the two regimes was studied, incorporating both viscous and thermal effects. Though not explored here, we note that the low frequency dispersion (\ref{ScalarAbsorb}) could be modeled with the tortuosity defined by (\ref{AlphaInvzz}) and (\ref{AlphaInvrr}) by the allowing the static permeability $K_0$ to be a complex function of frequency, such that
\begin{equation}\label{azzr}
\alpha_{zz}=\frac{\left(1+\frac{i\bar{\omega}\Gamma'}{c^2}\right)}{\left(1+\frac{\bar{\omega}^2\left(\Gamma'\right)^2}{c^4}\right)}.
\end{equation}
Alternatively, one could enforce this identification using the compressibility, $\beta$, inserted on the left-hand side of (\ref{azzr}).

Though the dynamics themselves cannot be related directly, one can still compare the absorption/amplification predicted the full tensor dispersion (\ref{DispersionOm}), the scalar dispersion (\ref{ScalarDisp}) with $\alpha^{-1}_{zz}$ given by (\ref{AlphaInvzz}), and the first-Born approximation (\ref{AbsFacc}). If we make the definition $\hat{\Gamma}\equiv \bar{\omega}_c c^2/\varphi\omega^2$, then in analogy with (\ref{Tfac})-(\ref{AbsFacc}), we have
\begin{equation}\label{TensAmp}
\text{Re}\left[\bar{k}_z\left(\frac{1}{\chi}-\chi\right)\right]=\frac{\bar{\omega}\omega^2\hat{\Gamma}}{k_z c^4}\cdot \Delta,
\end{equation}
where $\Delta\equiv (k_z c/\omega)^2+\delta$,
\begin{equation}\label{delta}
\delta \equiv \frac{\left(\varphi \omega c k_r\right)^2\left[\Omega_+^4+\bar{\omega}^2\left(\bar{\omega}_c^2+4\Omega^2\right)\right]}{\left(\Omega_-^4-\bar{\omega}^2\bar{\omega}_c^2\right)^2+4\bar{\omega}^2\bar{\omega}_c^2\Omega_+^4},
\end{equation}
and $\Omega_\pm^2\equiv \Omega^2\pm\bar{\omega}^2$. Figure \ref{fig:2} shows the difference between the $\Delta$ factors associated with scalar tortuosity, tensor tortuosity, and the first-Born approximation (the latter is obtained by setting $\hat{\Gamma}=\Gamma'$ and $\Delta=1$). One can immediately notice the breakdown of the scalar tortuosity approximation near $\bar{\omega}=0$, as mentioned above. Figure \ref{fig:2} also shows that the $\Delta$ factor predicted using the full tensor dispersion (\ref{DispersionOm}) is nearly two orders of magnitude larger than the first-Born approximation when $\bar{\omega}$ is close to zero. In this $\bar{\omega}$-near-zero region, the first order truncation of (\ref{Tasym}) is no longer appropriate, and one does not expect the first-Born approximation to accurately describe acoustic mode transmission, despite the low rotation rate and small disk thickness.

Ultimately, the level of description required to describe the superradiant amplification of acoustic OAM modes depends on the specific properties of the sound absorbing material, as well as on the parameter regime considered (details about experimental determination of the dissipative parameters can be found in \cite{Impedance,DissipationExp}). For rotation rates $\Omega$ that are large with respect to the convective frequency $\bar{\omega}\equiv\omega-m\Omega$, the tortuosity ceases to behave as a scalar, and the full tensor dispersion (\ref{DispersionOm}) should be used.

\section{Acknowledgements}

The author thanks Silke Weinfurtner, Bill Unruh, and Daniele Faccio, for stimulating discussions. This research was supported by the Royal Society Research Fellows Enhancement Award (2018, grant number RGF/EA/181015) held by Silke Weinfurtner at the University of Nottingham.

\end{document}